\documentclass[prl,amssymb,nofootinbib,superscriptaddress,twocolumn,longbibliography]{revtex4-2}
\pdfoutput=1
\usepackage[utf8]{inputenc}
\usepackage[english]{babel}

\usepackage{tikz}
\usetikzlibrary{shapes,backgrounds,fit,decorations.pathreplacing,arrows,decorations.markings}
\usepackage{verbatim}
\tikzstyle{vecArrow} = [thick, decoration={markings,mark=at position
   1 with {\arrow[semithick]{open triangle 60}}},
   double distance=1.4pt, shorten >= 5.5pt,
   preaction = {decorate},
   postaction = {draw,line width=1.4pt, white,shorten >= 4.5pt}]
\tikzstyle{innerWhite} = [semithick, white,line width=1.4pt, shorten >= 4.5pt]
\usepackage{bbm}
\usepackage{bm}
\usepackage{amsbsy}
\usepackage{amsthm}
\usepackage{amssymb}
\usepackage{amsfonts}
\usepackage{amsmath}
\usepackage{dsfont} % for symbols like the identity:\mathds{1}
\usepackage{graphicx}
\usepackage{epsfig}
\usepackage{epstopdf}
\usepackage{dsfont}
\usepackage{booktabs,multirow}
\usepackage{color}

\usepackage[colorlinks]{hyperref}
\makeatletter
\newcommand\org@hypertarget{}
\let\org@hypertarget\hypertarget
\renewcommand\hypertarget[2]{%
  \Hy@raisedlink{\org@hypertarget{#1}{}}#2%
  }
\makeatother
\usepackage[figure,table]{hypcap}
\usepackage{MnSymbol}
\usepackage{enumerate}%allows different styles of enumerate environment
\usepackage%[shortlables]
{enumitem}
\hypersetup{
	bookmarksnumbered,
	pdfstartview={FitH},
	citecolor={darkgreen},
	linkcolor={darkred},
	urlcolor={darkblue},
	pdfpagemode={UseOutlines}}
\definecolor{darkgreen}{RGB}{50,190,50}
\definecolor{darkblue}{RGB}{0,0,190}
\definecolor{darkred}{RGB}{238,0,0}
\definecolor{quantum}{RGB}{83,37,127}
\definecolor{quantumlight}{RGB}{169,146,191}
\usepackage{soul}

\newcommand{\ket}[1]{\ensuremath{\left|\right.\!{#1}\!\left.\right\rangle}}

\newcommand{\bra}[1]{\ensuremath{\left\langle\right.\!{#1}\!\left.\right|}}

\newcommand{\ketbra}[2]{\ensuremath{|{#1}\rangle\!\langle{#2}|}}

\renewcommand{\thesection}{\Roman{section}}
\renewcommand{\thesubsection}{\Roman{section}.\arabic{subsection}}

\begin{document}

\title{Average Equilibration Time for Gaussian Unitary Ensemble Hamiltonians}

\author{Emanuel Schwarzhans}
\email{emanuel.schwarzhans@um.edu.mt}
\affiliation{Department of Physics, University of Malta, Msida MSD 2080, Malta }
\affiliation{Atominstitut, Technische Universit{\"a}t Wien, 1020 Vienna, Austria}
\author{Alessandro Candeloro}
\email{alessandro.candeloro@unipa.it} 
\affiliation{School of Physics, Trinity College Dublin, Dublin 2, Ireland}
\affiliation{Università degli Studi di Palermo, Dipartimento di Fisica e Chimica - Emilio Segrè, via Archirafi 36, I-90123 Palermo, Italy}

\author{Felix~C.~Binder}
\email{felix.binder@tcd.ie} 
\affiliation{School of Physics, Trinity College Dublin, Dublin 2, Ireland}
\affiliation{Trinity Quantum Alliance, Unit 16, Trinity Technology and Enterprise Centre, Pearse Street, Dublin 2, Ireland}
\date{\today}

\author{Maximilian P. E. Lock}
\email{maximilian.paul.lock@tuwien.ac.at} 
\affiliation{Atominstitut, Technische Universit{\"a}t Wien, 1020 Vienna, Austria}
\affiliation{Institute for Quantum Optics and Quantum Information - IQOQI Vienna, Austrian Academy of Sciences, Boltzmanngasse 3, 1090 Vienna, Austria}
\date{\today}

\author{Pharnam Bakhshinezhad}
\email{faraj.bakhshinezhad@tuwien.ac.at} 
\affiliation{Atominstitut, Technische Universit{\"a}t Wien, 1020 Vienna, Austria}
\date{\today}
\begin{abstract}
Understanding equilibration times in closed quantum systems is essential for characterising their approach to equilibrium. Chaotic many-body systems are paradigmatic in this context: they are expected to thermalise according to the eigenstate thermalisation hypothesis and exhibit spectral properties well described by random matrix theory (RMT). While RMT successfully captures spectral correlations, its ability to provide quantitative predictions for equilibration timescales has remained largely unexplored. Here, we study equilibration within RMT using the framework of equilibration as dephasing, focusing on closed systems whose Hamiltonians are drawn from the Gaussian unitary ensemble~(GUE). We derive an analytical expression that approximates the average equilibration time of the GUE and show that it is independent of both the initial state and the choice of observable, a consequence of the rotational invariance of the GUE. Numerical simulations confirm our analytical expression and demonstrate that our approximation is in close agreement with the true average equilibration time of the GUE. We find that the equilibration time decreases with system size and vanishes in the thermodynamic limit. This unphysical result indicates that the true equilibration timescale of realistic chaotic many-body systems must be dominated by physical features not captured by random matrix ensembles -- the GUE in particular.
\end{abstract}

\maketitle

\section{Introduction}
Closed quantum systems evolve unitarily and therefore conserve the von Neumann entropy. Nevertheless, their dynamical expectation values typically relax towards stationary behaviour for long times. Over the past two decades, this apparent tension has been reconciled via an operational approach to equilibration, \emph{closed system equilibration of observables}: a system is said to equilibrate if, for most times, it is indistinguishable from its infinite time average (\emph{i.e. a time invariant}) state with respect some specific observable~\cite{reimann2008foundation}. In this formulation, equilibration is not a property of the state alone but of the pair $(\rho,A)$, that is, the state together with the observables used to probe it. Equilibration is thus fundamentally a statement about the long-time behaviour of expectation values of observables~\cite{Short_2012}.

It has been shown that for typical observables, equilibration can occur on very short timescales~\cite{goldstein2013,malabarba2014quantum,reimann2016typical}, and similarly, rapid equilibration has also been established for typical Hamiltonians~\cite{malabarba2014quantum}. At the same time, it is always possible to construct fine-tuned observables that equilibrate arbitrarily slowly, with equilibration times that can scale linearly with the Hilbert-space dimension \cite{goldstein2013,malabarba2014quantum}. 

For chaotic many-body systems, equilibration times have been linked to the dynamical manifestation of spectral correlations encoded in the spectral form factor~\cite{vallejo-fabila2025}. Depending on the regime considered, the equilibration time may decrease with system size or scale polynomially in system size. In particular, when equilibration is associated with features such as the correlation hole~\cite{torres-herrera2017} the relevant timescales are given by the Heisenberg time (\emph{i.e.} the inverse mean energy level spacing) which grows exponentially with system size~\cite{lezama2021equilibration}. Thus, while equilibration is generic, its timescale can vary dramatically depending on the structure of both the Hamiltonian and the observable.

To resolve this tension, recent work has shifted focus toward physically relevant observables and a more detailed understanding of the mechanisms underlying equilibration~\cite{garcia-pintos2017a,de2018equilibration}. In particular, equilibration can be understood as a dephasing process governed by the interference of energy gaps \cite{gogolin2016equilibration,de2018equilibration}. This perspective yields a refined notion of equilibration time that accounts for the relative relevance of different gaps, rather than relying solely on maximal spectral span. Conceptually, this approach is closely related to von Neumann’s quantum ergodic theorem~\cite{von_Neumann_2010,Goldstein_2010,Asadi_2015}, which established the equivalence between long-time averages and ensemble averages without addressing the associated timescales, and it has motivated modern investigations of equilibration based on structural properties of observables \cite{reimann2008foundation,garcia-pintos2017a}.

Chaotic many-body systems form a paradigmatic class of equilibrating systems. They are ubiquitous and typically thermalise, approaching steady states consistent with statistical mechanics, as formalised by the eigenstate thermalisation hypothesis (ETH) \cite{srednicki1994b,Rigol_2008_Thermalisation,Deutsch_2018_ETH}. Despite the complexity of their dynamics, the spectral statistics of chaotic systems in the bulk of the spectrum are remarkably well described by RMT~\cite{mehta1991,riddell2024}. This universality has made RMT a central tool for modeling chaotic quantum systems.

Although equilibration dynamics of chaotic systems and random matrices have been widely studied (see e.g. Ref.~\cite{vallejo-fabila2025}), it remains unclear to what extent universal spectral statistics alone determine the scaling of operational equilibration times~\cite{nickelsen2020}. In particular, a direct analytical derivation of a universal, system-size-dependent equilibration time from GUE spectral statistics is still lacking. Establishing such a connection would clarify whether equilibration times in chaotic quantum systems are fully determined by spectral correlations of random matrices, or whether additional information beyond RMT level statistics is required to accurately capture them.

In this paper, we address this question by analysing the equilibration time of closed quantum systems whose Hamiltonians are drawn from the GUE. Adopting the notion of equilibration time from dephasing, introduced in Ref.~\cite{de2018equilibration}, we derive an approximate analytical expression for the GUE-averaged equilibration time, which constitutes our main result:
\begin{align}\label{eq:Teq}
T_\mathrm{eq}^\mathrm{GUE}\approx \dfrac{c}{2\sigma}\,\dfrac{\pi}{\sqrt{2(N+1)}},
\end{align}
where $c$ is a constant, $\sigma$ the standard deviation of the GUE, and $N$ is the system Hilbert space dimension. Due to the rotational invariance of the GUE~\cite{mehta1991}, this average equilibration time is independent of the observable and the initial state and vanishes in the limit of large Hilbert-space dimension $N$. We further provide numerical evidence for its validity by explicitly simulating the dynamics of GUE sampled Hamiltonians and computing their average equilibration time using a proxy quantity. Evaluating the proportionality constant~$c$, we find that the numerically evaluated~$c$ is close to its analytical prediction (App.~\ref{app:InverseOfAverage}~and~\ref{app:unimod-constant}).

Our results show that, in the absence of any locality structure, the spectrum alone determines equilibration timescales and RMT spectral statistics imply average equilibration times that vanish in the large-system limit. Since physical many-body systems cannot equilibrate instantaneously this demonstrates that RMT statistics are insufficient to capture physically meaningful equilibration timescales. This indicates that features absent from RMT, such as the structure of the spectral edges and deviations from perfect rotational invariance in the bulk, must play a crucial role in setting finite equilibration times in large physical systems. Consequently, care must be taken when using RMT as a dynamical model for chaos: while it excels at describing spectral universality, it can be misleading when applied to equilibration timescales.

\section{Equilibration time scales of closed systems}
To understand how closed quantum systems 
equilibrate, we follow the framework of  
\emph{average equilibration of observables}~\cite{garcia-pintos2017a,Short_2012,linden2009quantum,reimann2008foundation}.  
A closed system is said to equilibrate with respect to an observable \(A\) if, for most times \(t\), the expectation value 
\(\langle A\rangle_t=\mathrm{Tr}[A\rho(t)]\) is practically indistinguishable from that of its \emph{equilibrium state}.  
Furthermore, if the expectation value of $A$ equilibrates on average, it will equilibrate towards its infinite time average $\langle A \rangle_\mathrm{eq} = \mathrm{Tr}[A\rho_{\mathrm{eq}}(t)]$, where $\rho_{\mathrm{eq}}(t)$ is given by
\begin{align}\label{eq:rho_eq}
\rho_\mathrm{eq}
:= \lim_{T\rightarrow \infty} \frac{1}{T}\int_0^T \rho(t)\,dt,
\end{align}
which is the state that maximises the entropy given all conserved quantities~\cite{gogolin2016equilibration}. 
To quantify equilibration, we consider the \emph{time signal}
\begin{align}
g(t) := \frac{1}{\Delta_A}\mathrm{Tr}\big[A(\rho(t)-\rho_\mathrm{eq})\big],
\end{align}
where $\Delta_A=a_\mathrm{max} -a_\mathrm{min   }$ normalises for the difference between maximal and minimal eigenvalue of $A$. Then \(|g(t)|^2\) quantifies the squared deviation of the expectation value of $A$ from its equilibrium value at time $t$.
Equilibration requires that this deviation be small for almost all times.  
Consequently, its infinite time-average must be close to zero,
\begin{align}
\big\langle |g(t)|^2 \big\rangle_\infty :=
\lim_{T\rightarrow\infty}\frac{1}{T}\int_0^T |g(t)|^2 dt
\approx 0.
\end{align}
Furthermore, fluctuations away from the equilibrium value of $A$ at any point in time are typically exponentially suppressed in the system's dimension~\cite{reimann2008foundation},
\begin{align}\label{eq:reimann}
   \mathrm{Prob}\!\left[
   \Big|\mathrm{Tr }[A\rho(t)]-\mathrm{Tr}[A\rho_\mathrm{eq}]\Big|
   \geq \frac{\Delta_A}{d_\mathrm{eff}^{1/3}}
   \right]
   < \frac{1}{d_\mathrm{eff}^{1/3}},
\end{align}
where $d_\mathrm{eff} := 1/\mathrm{Tr}(\rho_\mathrm{eq}^2)$ is the \emph{effective dimension} -- sometimes called inverse participation ratio -- and  
$\Delta_A$ is the spectral range of $A$ defined as the difference between the largest and smallest eigenvalues of $A$ whose spectrum is assumed to be bounded.  
Thus, when $d_\mathrm{eff}$ is large, fluctuations of the equilibrium value are strongly suppressed as the effective dimension grows, and $A$ is effectively equilibrated
\begin{align}
A\text{ equilibrated} \quad \Longleftrightarrow \quad 
\left\langle |g(t)|^2 \right\rangle_\infty \approx 0.
\end{align}
Typically, macroscopic quantum systems explore a significant portion of the Hilbert space during their evolution, which is characterised by a large effective dimension. Such systems equilibrate for a broad class of observables~\cite{reimann2008foundation,short2011}. Equipped with this notion of equilibration, we can now look at the timescales involved. 

To do so, we will consider the framework of \emph{equilibration as dephasing} which was developed in Refs.~\cite{gogolin2016equilibration,de2018equilibration}. We will outline the basic idea of this framework following Ref.~\cite{de2018equilibration}: Consider a system Hamiltonian $H$ which can be decomposed as $H=\sum_k E_k \ketbra{E_k}{E_k}$, where $E_k$ are its energies and $\{\ket{E_k}\}_k$ are energy eigenstates. Assume that the system starts in a pure initial state $\ket{\psi(0)}=\sum_k c_k \ket{E_k}$. Its time-evolved state is then given by 
\begin{align}\label{eq:rho0}
    \ket{\psi(t)}=\sum_k c_k e^{-i E_k t} \ket{E_k}.
\end{align}
If it equilibrates, the system's equilibrium state is given by the infinite-time average state~\eqref{eq:rho_eq}
\begin{align}
    \rho_\mathrm{eq}=\sum_k |c_k|^2\ketbra{E_k}{E_k}.
\end{align}
In this case, the time signal of $A$ is given by
\begin{align}
    &g(t):=\frac{1}{\Delta_A}(\langle\psi(t)|A| \psi(t)\rangle-\operatorname{Tr}(A \rho_\mathrm{eq}))\\&=\frac{1}{\Delta_A} \sum_{i \neq j}\left(c_j^* A_{j i} c_i\right) \mathrm{e}^{-\mathrm{i}\left(E_i-E_j\right) t}=\sum_{ij} \nu_{ij} \mathrm{e}^{\mathrm{i} G_{ij} t}\,,
    \label{eq:signalg(t)}
\end{align}
where $A_{ij}=\bra{E_i}A\ket{E_j}$ are the matrix elements of $A$ in the energy eigenbasis, $\nu_{ij}=\tfrac{c_j^* A_{j i} c_i}{\Delta_A}$, and $G_{ij}=E_j-E_i$. Thus, whether a system equilibrates depends on its initial state $\rho(0)$, the observable $A$ under consideration, and the energy gap structure of $H$, which determines $G_{ij}$. 

Rewriting $\nu_{ij}$ as $\nu_{ij}=|\nu_{ij}|\,e^{i \theta_{ij}}$ which can be thought of as vectors on the complex plane, each evolving corresponding to individual Fourier components of the Hamiltonian~\cite{de2018equilibration}. These rotate around the complex plane with different velocities, which are determined by the gaps $G_{ij}$. For an initial state far from equilibrium (for some observable $A$) the vectors $\nu_{ij}$ are approximately phase-aligned. Equilibration happens when these dephase completely -- that is, when their sum equals (or is close to) the origin, which implies that~\eqref{eq:signalg(t)} vanishes. Thus, to determine the timescales at which equilibrium is approached, one has to take into account that each $\nu_{ij}$ contributes with a different strength to $g(t)$ due to their differing amplitudes. This is captured by the so-called \emph{gap relevances}~\cite{de2018equilibration}, 
\begin{align}
    q_{ij}=\dfrac{|\nu_{ij}|^2}{\sum_{k,l} |\nu_{kl}|^2}.
\end{align}
The time it takes for the system to equilibrate is therefore determined by how long it takes phases of $\nu_{ij}$ to assume a configuration where they collectively sum to zero. If all vectors evolve with a similar velocity, corresponding to a small variance in the distribution of energy gaps, equilibration occurs slowly and equilibration is delayed. Faster equilibration happens, when the variance is large, leading to more rapid dephasing. The rate at which this dephasing happens is governed by the dispersion of gaps in the complex plane $\sigma_G$, given by the variance of energy gaps:
\begin{align}
    \sigma_G^2=\sum_{ij}q_{ij} (G_{ij}-\mu_G)^2=\sum_{ij}q_{ij} G_{ij}^2,
    \label{eq:gapdisp}
\end{align}
where $\mu_G$ is the mean gap, which vanishes here due to $G_{ij}=-G_{ji}\;\forall i,j$. To understand how the equilibration time can be estimated from this description, consider the scenario where all gap relevances are identical and the initial state starts with all phases perfectly aligned. In this case the time it takes for the system to equilibrate can be lower bounded by the time it takes for the fastest and slowest rate to differ by $\pi$, \emph{i.e.}, $t (G_\mathrm{max} -G_\mathrm{min})=\pi$, at which point they cancel each other out. This, in turn, leads to an equilibration time $T_\mathrm{eq}\sim \pi/(G_\mathrm{max} -G_\mathrm{min})$. Since the extreme gaps might not be the most relevant contributions to~\eqref{eq:signalg(t)}, a more representative estimate is obtained by weighting gaps by their relevance. For a unimodal distribution, for instance, in which the most relevant gaps are concentrated around a single peak and the tails contribute negligibly, the equilibration time is well-estimated by the width of the relevance-weighted gap distribution $\sigma_G$. More generally, replacing $G_\mathrm{max} -G_\mathrm{min}$ with $2\sigma_G$ gives a more accurate estimate of the equilibration time,
\begin{align}\label{eq:EquilibrationTimeHeuristic}
    T_\mathrm{eq}\approx \dfrac{\pi}{2\sigma_G},
\end{align}
regardless of the underlying distribution. In App.~\ref{app:unimod-constant}, we demonstrate numerically that the GUE has a unimodal gap distribution, which further justifies this estimate. In the remainder of this work, we apply this definition of equilibration time to random matrix ensembles in order to analyse its ensemble-averaged behaviour.

\section{Quantum Chaos and Random Matrix Theory}

RMT provides a framework for modeling the spectral statistics of chaotic quantum systems. According to the Bohigas–Giannoni–Schmit (BGS) conjecture \cite{bohigas1984}, the energy-level statistics of quantum systems with a classically chaotic counterpart follow those predicted by RMT. 
These statistics are universal in the bulk of the spectrum and are characterised by Wigner–Dyson level spacing distributions, featuring level repulsion. 
By contrast, integrable systems obey the Berry–Tabor conjecture \cite{berry1997} and display Poissonian level statistics, with uncorrelated gaps. 
Thus, the presence of level repulsion is a strong, though not sufficient, indicator of quantum chaos \cite{gu2024simulating}. 
RMT is therefore ideally suited to study generic properties of chaotic systems when edge effects and system-specific symmetries are not dominant.

To model chaotic Hamiltonians, we draw them from the GUE, which represents the universal class of complex Hermitian matrices without time-reversal symmetry \cite{mehta1991}. 
The GUE is invariant under unitary transformations, $P(H)=P(U^\dagger H U)$ for any unitary $U$. As a consequence, the ensemble-averaged dynamics depends only on the spectrum and is independent of the eigenbasis, making it impossible to single out a preferred tensor-product structure and rendering any notion of locality meaningless.

For any Hilbert space of dimension $N$, the GUE consists of $N \times N$ Hermitian matrices $H$ whose entries $H_{ij}$ are independent Gaussian variables, 
\begin{align}\label{eq:GUEelements}
    P(H_{ij}) = \frac{1}{\sigma \sqrt{2\pi}} \exp\!\left[-\frac{H_{ij}^2}{2\sigma^2}\right],
\end{align}
with the constraint $H_{ji}=H_{ij}^*$ for $i \neq j$. 
The joint probability density for obtaining a specific $H$ can be written in terms of the eigenvalues $\{E_i\}$ as~\cite{mehta1991}
\begin{align}
    P(H)\equiv P(\{E_i\}) = \frac{1}{\mathcal{Z}} 
    \prod_{i<j}^{N-1} |E_i -     E_j|^2 
    e^{-\frac{1}{2\sigma^2}\sum_{i=0}^{N-1} E_i^2},
\end{align}
where $\mathcal{Z}=(2\pi)^{N/2} \sigma^{2N} \prod_{i=1}^N i!$ is the partition function and the factor $\prod_{i<j}|E_i - E_j|^2$ encodes level repulsion. 
We will use GUE sampling throughout this work to generate Hamiltonians representative of chaotic systems and to study their equilibration properties.

\section{Average equilibration time for the GUE}
Consider a Hamiltonian $H$ drawn from the GUE according to 
\eqref{eq:GUEelements}, with zero mean and standard deviation~$\sigma$.  
Starting from an arbitrary initial state $\ket{\psi(0)}$ and considering an arbitrary observable $A$, we find that -- owing to the unitary 
invariance of the GUE -- the average energy-gap dispersion~\eqref{eq:gapdisp} 
is independent of both the initial state and the choice of observable, depending 
solely on the variance of the energy gaps, \emph{i.e.} 
\begin{align}
& \left\langle\sigma_G^2\right\rangle_{G U E}=\int_{-\infty}^\infty ... \int_{-\infty}^\infty \Pi_n d E_n \sigma_G^2 P\left(E_1, E_2,..., E_N\right)\\&
=\iint_{-\infty}^\infty d E_i d E_j\left(E_j-E_i\right)^2 P\left(E_i, E_j\right),
\end{align}
where the second equality comes from the fact that $\sigma_G$ only depends on pairs of $E_i$ and $E_j$ throught the difference $|E_j-E_i|$, thus allowing us to integrate the probability distribution over the remaining energies $E_k$ for $k\neq i,j$ leaving us with the probability distribution for the pair $E_i$ and $E_j$ (see App.~\ref{app:Bounding the equilibration time of Gaussian Unitary Ensembles} for further details). 

We can evaluate this integral analytically (see App.~\ref{app:Bounding the equilibration time of Gaussian Unitary Ensembles}), yielding
\begin{align}\label{eq:gapdispersion}
    \left\langle\sigma_G^2\right\rangle_{G U E}=2\sigma^2(N+1).
\end{align}
Since the equilibration time defined in~\eqref{eq:EquilibrationTimeHeuristic} is that of a single instance, to get the average equilibration time of GUE Hamiltonians we have to take the ensemble average $\langle T_\mathrm{eq}\rangle_\mathrm{GUE}$. To express this in terms of the average gap dispersion we use that for a random variable $X$ the following relation holds
\begin{align}\label{eq:deltamethod}
        \left\langle\frac{1}{\sqrt{X}}\right\rangle \approx \frac{1}{\sqrt{\left\langle X\right\rangle_\mathrm{GUE}}}\left(1+\frac{3}{8} \frac{\operatorname{Var}\left(X\right)}{\left\langle X\right\rangle_\mathrm{GUE}^2}\right).
\end{align}
where $X=\sigma_G^2$ in our case. In App.~\ref{app:InverseOfAverage} we show that the relative variance of the energy gap dispersion is bounded as
\begin{align}
    0\leq\dfrac{\operatorname{Var}(\sigma_G^2)}{\langle \sigma_G^2\rangle_\mathrm{GUE}^2}\leq 1.
\end{align}
This allows us to substitute~\eqref{eq:gapdispersion} in~\eqref{eq:EquilibrationTimeHeuristic} to get the average equilibration time of $N$-dimensional GUE sampled Hamiltonians as reported the main result~\eqref{eq:Teq}.

\begin{figure}[t]
    \centering
    \includegraphics[width=1\linewidth]{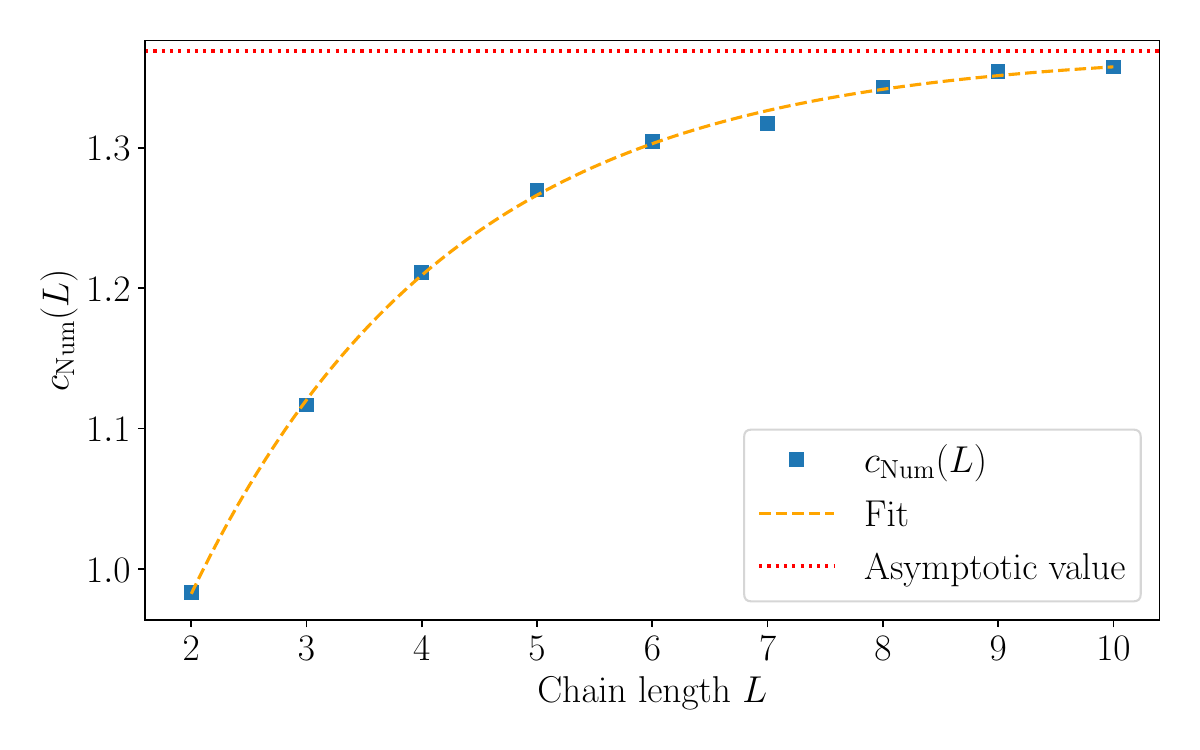}
    \caption{Plot of the average proportionality constant $c_\mathrm{Num}(L)$ as a function of the number of qubits $L$ (where $N=2^L$) for randomly sampled GUE Hamiltonian. Blue squares: $c_\mathrm{Num}(L)$ computed using~\eqref{eq:unimodal_c_fit}, where $t_\mathrm{fc}$ is obtained by averaging $n=1000$ randomly sampled GUE matrices for each value of $L$. Orange dashed line: fit of the raw data with respect to shifted exponential function $A e^{-B*L} + c_\mathrm{Num}$. Red dotted line: apparent asymptotic value $c_\mathrm{Num}$ for large $L$, obtained from the fit. The variance is set to $\sigma = 1/\sqrt{N}$. Further details regarding this choice and the fit can be found in App.~\ref{app:unimod-constant}.}
    \label{fig:fit_c_vs_L}
\end{figure}

To verify this relation, we perform a numerical simulation of the dynamics of GUE sampled Hamiltonians. As mentioned above, due to the rotational invariance of the GUE, the average equilibration time is independent of the choice of observable $A$. We exploit this by choosing $N=2^L$, allowing us to interpret $H_{\text{GUE}}$ as the Hamiltonian of $L$ spin $1/2$ systems with all-to-all interactions. 
As observable we choose the bulk magnetisation, given by the sum of local spin components $A=\sum_{i=1}^{L} \sigma_z^{(i)}$. As a proxy for the equilibration time we calculate the ``first crossing time" $t_\mathrm{fc}$, that is the time when $|g(t)|^2\leq\langle |g(t)|^2\rangle_\infty$ for the first time.  For sufficiently large $N$, this proxy is well motivated, as fluctuations away from equilibrium are strongly suppressed (see~\eqref{eq:reimann} and App.~\ref{app:unimod-constant}). With this, we can express the (numerically evaluated) proportionality constant as
\begin{equation}
    c_\mathrm{Num} (N)\approx t_\mathrm{fc}\dfrac{2\sqrt{2\sigma^2(N+1)}}{\pi}\,.
    \label{eq:unimodal_c_fit}
\end{equation}
Fig.~\ref{fig:fit_c_vs_L} shows the numerical evaluation of this constant for up to $N=2^{10}$. Extrapolating these data points suggests that $c_\mathrm{Num}$ approaches a finite asymptotic value for large~$N$ given by
\begin{align}
    c_\mathrm{Num}(N)\xrightarrow{N\rightarrow\infty}1.368 \pm 0.004,
\end{align}
which is very close to the analytical value of $c$
\begin{align}    
c\approx\left( 1+\tfrac{3}{8}\tfrac{\operatorname{Var}\left(X\right)}{\left\langle X\right\rangle_\mathrm{GUE}^2}\right)\xrightarrow{N\rightarrow\infty}1.375
 \end{align}
 derived in App.~\ref{app:InverseOfAverage}. This implies that~\eqref{eq:Teq} approximates the true equilibration time well for large $N$. Additionally, this demonstrates that $c$ is indeed independent of system size, meaning that it does not influence the scaling of $T_\mathrm{eq}^\mathrm{GUE}$ with $N$ for large systems. Further details on the numerical analysis are provided in App.~\ref{app:unimod-constant}. 

 Since $c$ indeed is constant for large $N$, we can conclude that~\eqref{eq:Teq} shows that the equilibration time $T_\mathrm{eq}^\mathrm{GUE}$ vanishes as $N \to \infty$, indicating that bulk spectral contributions (where RMT applies) become negligible in the thermodynamic limit. Consequently, since realistic systems do not equilibrate instantaneously, deviations from RMT and spectral edge effects must be the dominant contribution to the equilibration time in realistic chaotic many-body systems and RMT contributions become irrelevant. 

\section{Discussion and Conclusion}

In this work, we analytically derived an approximate expression for the typical equilibration time of chaotic quantum systems modelled by RMT. In particular, the expression applies to closed systems whose Hamiltonians are sampled from the GUE~\eqref{eq:Teq}. We found that the resulting equilibration time is, in fact, independent of the initial state and choice of observable. This can be understood as a consequence of the rotational invariance of the GUE. We further provided evidence showing that~\eqref{eq:Teq} closely approximates the true average equilibration time of the GUE by sampling Hamiltonians from the GUE and numerically simulating their time evolution and equilibration properties. In doing so, we also demonstrated that the notion of equilibration as dephasing introduced in Ref.~\cite{de2018equilibration} applies consistently to systems that can be reliably modelled by RMT. 

A key result of our analysis is that the equilibration time~\eqref{eq:Teq} decreases with increasing Hilbert-space dimension and vanishes in the limit of large system size. Since instantaneous equilibration does not occur in physical systems, this behaviour indicates that modelling many-body dynamics using pure random-matrix ensembles cannot capture all features relevant for realistic equilibration. RMT is, after all, only an effective description of the universal spectral statistics of complex systems, expected to apply primarily in the bulk of many-body spectra and even there only approximately. Our results therefore suggest that the contribution to equilibration coming from the RMT-like part of the spectrum becomes negligible for large systems, and that finite equilibration times observed in realistic systems must be determined by deviations from RMT.

Such deviations can arise in several ways. As noted above, the edges of the energy spectrum of many-body systems are known to deviate from the spectral distributions predicted by RMT \cite{bohigas1984}. Moreover, even within the bulk, physical constraints can lead to spectral properties that differ from those of the GUE \cite{bohigas1971,feng2024}. Consistent with this picture, studies of concrete interacting models have found equilibration times that scale as a power law or even exponentially with system size \cite{lezama2021equilibration}, in stark contrast to the vanishing timescale obtained for GUE Hamiltonians. More generally, investigations of thermalisation dynamics in chaotic many-body systems have shown that dynamical timescales can depend sensitively on microscopic structure and may not follow the behaviour expected from pure random-matrix models \cite{vallejo-fabila2025}. In addition, non-ergodic features such as many-body scars can lead to deviations from random-matrix behaviour, and can affect equilibration dynamics \cite{turner2018,serbyn2021}.

Furthermore, sampling from the GUE results in a dense Hermitian matrix with i.i.d. entries that are all drawn from the same Gaussian distribution. Such a matrix couples every degree of freedom to every other, leading to an all-to-all connected system. This is a direct reflection of the fact that the GUE is invariant under unitary rotations, meaning that there exists no preferred tensor-product structure, rendering any notion of locality meaningless. It is consistent with the study of equilibration time according to the interconnectedness of a many-body Hamiltonian~\cite{nickelsen2020}. This contrasts with typical physical scenarios that feature local interactions, where equilibration is often limited by propagation constraints such as Lieb-Robinson-like bounds~\cite{lieb2004}. This lack of locality provides a natural explanation for the qualitative difference between the vanishing equilibration times found here and the finite, system-size-dependent timescales observed in real-world based models.

Our findings also contrast with the fast-scrambling conjecture, which predicts that thermalisation times in chaotic, all-to-all coupled systems scale logarithmically with the number of degrees of freedom, as diagnosed for instance via out-of-time-order correlators \cite{sekino2008fast,belyansky2020minimal}, indicating that locality may not be the only reason for finite equilibration times. The vanishing average equilibration time likely indicates a failure of RMT to capture effects that go beyond locality alone. 

In conclusion, our results illustrate that, while RMT provides a powerful framework for describing spectral properties of chaotic systems, it fails to capture physically meaningful equilibration times. The vanishing equilibration time of GUE Hamiltonians implies that finite equilibration times in real chaotic systems must originate from precisely those features -- such as spectral edges, structured deviations in the bulk, or other effects -- that lie beyond the scope of RMT.

\raggedbottom

\section*{Acknowledgements}

The authors want to thank Yuri Minoguchi for fruitful discussions.
E.S. acknowledges the support of the European Union via the Quantum Flagship project ASPECTS (Grant Agreement No.\ 101080167).
A.C. acknowledges support from the "Italian National Quantum Science and Technology Institute (NQSTI)" (PE0000023) - SPOKE 2 through project ASpEQC. P.B. acknowledges funding from the European Research Council (Consolidator grant ‘Cocoquest’ 101043705) and financial support from the Austrian Federal Ministry of Education, Science and Research via the Austrian Research Promotion Agency (FFG) through the project FO999914030 (MUSIQ) funded by the European Union – NextGenerationEU. Further financial support was provided by the Austrian Science Fund (FWF) through the Stand-Alone grant P 36633-N, as well as 10.55776/I6949, and 10.55776/COE1. F.C.B. acknowledges support by Taighde Éireann - Research Ireland under grant number IRCLA/2022/3922.

%%%%%%%%%%%%%%%%%%%%%%%%%%%%%%%%%%
\bibliographystyle{apsrev4-1fixed_with_article_titles_full_names.bst}
\bibliography{bibfile.bib}
\onecolumngrid
\appendix

\renewcommand{\thesubsection}{\thesection.\arabic{subsection}}
\renewcommand{\thesubsubsection}{\thesubsection.\arabic{subsubsection}}

%%%%%%%%%%%%%%%%%%%%%%%%%%%%%%%%%%%%%%%%%%
\section*{Appendices}

\section{Bounding the equilibration time of Gaussian Unitary Ensembles}\label{app:Bounding the equilibration time of Gaussian Unitary Ensembles}
In this appendix, our aim is to calculate the equilibration time of Hamiltonians sampled from the Gaussian Unitary Ensemble (GUE) using the definition of equilibration of~\eqref{eq:EquilibrationTimeHeuristic}: 
\begin{align}\label{eq:equilibrationtime_non_averaged}
    T_{eq}\approx \dfrac{\pi}{2\sigma_G}=\dfrac{\pi}{2\sqrt{\sum_\alpha q_\alpha (G_\alpha - \mu_G)^2}}\,,
\end{align}
where $q_\alpha = \frac{\vert \nu_\alpha \vert^2}{\sum_\beta \vert \nu_\beta\vert^2}$ are called relevance (of the gap $\alpha$) and $\nu_\alpha=\nu_{ij}=\tfrac{1}{\Delta_A}c_i A_{ij} c_j^*$, with the initial state given by $\ket{\psi_0}=\sum_i c_i \ket{e_i}$ where the normalisation factor $\Delta_A=a_\mathrm{max}-a_\mathrm{min}$ accounts for the range of possible outcomes of the observable $A$ on the total system. $G_\alpha=G_{ij}=E_j-E_i$ are the energy gaps of the Hamiltonian. 
We assume that the Hamiltonian is sampled from a GUE such that each matrix element is selected from a Gaussian distribution with mean value $0$ and variance $\sigma$. First, we note that the $\nu_\alpha$ are underdetermined, as we made no assumptions on the initial state $\ket{\psi_0}$ or the observable $A$. However, using GUE Hamiltonians allows us to circumvent this underdetermination.

First, note that, for the GUE, we can simplify the calculation of the average equilibration time by approximating the average of the inverse square root, by the inverse square root of the average, \emph{i.e.}
\begin{align}
    T^\mathrm{GUE}_\mathrm{eq}:=\langle T_\mathrm{eq}\rangle_{GUE} = \left\langle\dfrac{\pi}{2\sigma_G}\right\rangle_\mathrm{GUE} \approx \dfrac{c\pi}{\sqrt{\langle\sigma_G^2\rangle_{GUE}}},
\end{align}
where $c$ is a proportionality constant. In App.~\ref{app:InverseOfAverage} we show that this hold.

The probability of picking a specific Hamiltonian $H$ in the region $dH=\prod_{i,j}dH_{ij}$  from the GUE with a standard deviation $\sigma$ is given by 
\begin{align}
    P(H)=P(E_1,E_2,...,E_N)=\dfrac{1}{\mathcal{Z}}e^{-\frac{1}{2\sigma^2}tr(H^2)}dH=\dfrac{1}{\mathcal{Z}}e^{-\frac{1}{2\sigma^2}\sum_i E_i^2}\prod_{i\neq j}|E_j-E_i|^2dE_1dE_2...dE_N\,,
\end{align}
where $\mathcal{Z}=(2\pi)^{N/2} \sigma^{2N} \prod_{i=1}^N i!$ is the partition function. For now, however, the specific form of this distribution is not our primary concern, what matters is that it depends only on the eigenvalues of the Hamiltonian and that its statistics are invariant under unitary rotations of the eigenvectors
\begin{align}
c_iA_{ij}c_j^*\stackrel{Pr.}{=}\tilde{c}_i\tilde{A}_{ij}\tilde{c}_j^*= \Big(\langle\psi_0|U\Big)\ketbra{e_i}{e_i}\Big(UAU^\dagger\Big) \ketbra{e_j}{e_j}\Big(U|\psi_o\rangle\Big)\,,
\end{align}
where $\stackrel{Pr.}{=}$ means "equal in probability".
So picking a Hamiltonian with certain eigenvalues $E_1,E_2...$ is equally probable for all eigenvectors. This allows us to calculate the ensemble average of $\sigma^2_G$ over the GUE by using its probability distribution $P(E_1,E_2...)$
\begin{align}
&\langle \sigma_G^2 \rangle_{GUE}=\underbrace{\int_{-\infty}^\infty dE_1\int_{-\infty}^\infty dE_2 ... \int_{-\infty}^\infty dE_N }_{=\int_{-\infty}^\infty...\int_{-\infty}^\infty \prod_n dE_n} \;\sigma_G^2 \, P(E_1,E_2,...,E_N)\\&
=\int_{-\infty}^\infty dE_1\int_{-\infty}^\infty dE_2 ... \int_{-\infty}^\infty dE_N  \sum_{ij}q_{ij} \, (E_j-E_i)^2 \, P(E_1,E_2,...,E_N).
\end{align}
Now since $q_\alpha=\frac{|\nu_\alpha|^2}{\sum_\beta |\nu_\beta|^2}$ and $\nu_\alpha=\nu_{ij}=c_i A_{ij} c_j^*=\langle\psi_0\ketbra{e_i}{e_i}A \ketbra{e_j}{e_j}\psi_0\rangle$ where $\psi_0$ is the initial state of the system, $A$ the observable and $\ket{e_i}$ energy eigenstates, the distribution relevances $q_\alpha$ for GUE does not depend on the eigenvalues $E_1,E_2,...,E_N$. This allows us to make the following simplifications
\begin{align}
& \int_{-\infty}^\infty dE_1\int_{-\infty}^\infty dE_2 ... \int_{-\infty}^\infty dE_N \sum_{ij} \, q_{ij} \, (E_j-E_i)^2 \, P(E_1,E_2,...,E_N)\\&
= \sum_{ij}q_{ij} \int_{-\infty}^\infty dE_idE_j (E_j-E_i)^2\underbrace{\int_{-\infty}^\infty \prod_{k\neq i,j} \, dE_k \, P(E_1,E_2,...,E_N)}_{P(E_i,E_j)}\\&
=\underbrace{\sum_{ij}q_{ij}}_{=1}\underbrace{\iint_{-\infty}^\infty dE_idE_j \, (E_j-E_i)^2\, P(E_i,E_j)}_{\mathrm{indep.\; of }\;i,j}\,.
\end{align}
We can thus write the average energy gap dispersion as
\begin{align}\label{eq:gapdispresionsimple}
    \langle\sigma_G^2\rangle_\mathrm{GUE}=\iint_{-\infty}^\infty dE_idE_j\; (E_j-E_i)^2P(E_i,E_j)\,,
\end{align}
where the last equality comes from the fact that the joint probability distributions $P(E_i,E_j)$ is the same for all $i,j$ (see page 60 of \cite{livan2018}).

This leaves us with the task of evaluating the integral in the denominator, which can be done explicitly. As a first step note that $P(E_i,E_j)$ has been calculated explicitly (see page 60 of \cite{livan2018}) to be 
\begin{align}
\label{eq:jointprob}
    P(E_i,E_j)= \dfrac{1}{N(N-1)}(K_N(E_i,E_i)K_N(E_j,E_j)-K^2_N(E_i,E_j))\,,
\end{align}
where the kernel $K_N(x,y)$ is given as 
\begin{align}
\label{eq:kernel}
    K_N(x,y)=e^{-\frac{1}{2}(V(x)+V(y))}\sum_{j=0}^{N-1}\pi_j(x) \, \pi_j(y),
\end{align}
where $V(x)$ for the GUE with variance $\sigma$ is given by $V(x)=x^2/2\sigma^2$ and equally for $y$, and $\pi_j(x)=\dfrac{H_j(x/\sqrt{2}\sigma)}{\sqrt{\sqrt{2\pi}2^j j!\sigma}}$ where $H_j(x)$ are Hermite polynomials of order $j$. To get orthonormality of the $\pi_j(x)$ here we calculate the normalisation factor by using the fact that $\int_{-\infty}^\infty dx H_k(x)H_{l}(x) e^{-x^2}=\sqrt{\pi}2^k k! \delta_{k l}$ replacing $x$ with $\tfrac{x}{\sqrt{2} \sigma}$, where substitution gives $\int_{-\infty}^\infty dx H_k(\tfrac{x}{\sqrt{2} \sigma})H_{l}(\tfrac{x}{\sqrt{2} \sigma}) e^{\tfrac{x^2}{2 \sigma^2}}=\sqrt{2\pi} \, 2^k k! \, \sigma \, \delta_{k l}$, and then dividing by the square root of the factor on the right side for normalisation.
Inserting~\eqref{eq:kernel} in~\eqref{eq:gapdispresionsimple} we get
\begin{align}\label{eq:full_integral}
&\langle \sigma_G^2\rangle_{GUE} =  \iint_{-\infty}^\infty dE_idE_j\;\dfrac{(E_i^2 + E_j^2 - 2 E_i E_j)}{N(N-1)}e^{-\frac{1}{2\sigma^2}(E_i^2+E_j^2)}\left(\sum_{k,l=0}^{N-1}\pi_k^2(E_i)\pi_{l}^2(E_j)-\sum_{k,l=0}^{N-1}\pi_k(E_i)\pi_{l}(E_i)\pi_k(E_j)\pi_{l}(E_j)\right) \nonumber\\&
= \dfrac{1}{N(N-1)}\iint_{-\infty}^{\infty}dE_idE_j\, e^{-\frac{1}{2\sigma^2}(E_i^2+E_j^2)}\left((E_i^2 + E_j^2)\sum_{k,l=0}^{N-1}\pi_k^2(E_i)\pi_{l}^2(E_j)- (E_i^2 + E_j^2)\sum_{k,l=0}^{N-1}\pi_k(E_i)\pi_{l}(E_i)\pi_k(E_j)\pi_{l}(E_j) \right. \nonumber\\  & \left.
-2 E_i E_j\sum_{k,l=0}^{N-1}\pi_k^2(E_i)\pi_{l}^2(E_j)+2 E_i E_j\sum_{k,l=0}^{N-1}\pi_k(E_i)\pi_{l}(E_i)\pi_k(E_j)\pi_{l}(E_j)\right)\,.
\end{align}
We divide this calculation into four parts, two for terms with prefactor $E_i^2+E_j^2$ and two with prefactor $2E_iE_j$, and evaluate each one separately. Before we evaluate this integral it is instructive to recall the following relation for (the probabilist's) Hermite polynomials:

\begin{align}
x H_n(x)=\frac{1}{2} H_{n+1}(x)+n H_{n-1}(x).
\end{align}

For our modified polynomial we have
\begin{align}
\pi_k(x)
= \frac{H_k\!\left(x / \sqrt{2}\sigma\right)}{\sqrt{\sqrt{2\pi}\,2^k k!\,\sigma}}
= \frac{H_k\!\left(x / \sqrt{2}\sigma\right)}{N_k},
\end{align}
where
\begin{align}
N_k=\sqrt{\sqrt{2\pi}\,2^k k!\,\sigma}.
\end{align}
We now want to evaluate
\begin{align}
x \pi_k(x)
= \frac{x H_k\!\left(\frac{x}{\sqrt{2}\sigma}\right)}{N_k}.
\end{align}
Substituting $y=\frac{x}{\sqrt{2}\sigma}$ gives
\begin{align}
\frac{\sqrt{2}\sigma}{N_k} \, y H_n(y)
= \frac{\sqrt{2}\sigma}{N_k}
\left(
\frac{1}{2} H_{k+1}(y)+k H_{k-1}(y)
\right).
\end{align}

Now notice that
\begin{align}
\frac{N_{k+1}}{N_k}=\sqrt{2(k+1)}, \quad \frac{N_{k-1}}{N_k}=\frac{1}{\sqrt{2 k}} .
\end{align}

Inserting this into the expression above gives
\begin{align}
\frac{\sqrt{2}\sigma}{N_k} y H_k(y)
&=\sigma\left(
\sqrt{k+1}\frac{H_{k+1}(y)}{N_{k+1}}
+
\sqrt{k}\frac{H_{k-1}(y)}{N_{k-1}}
\right)
\\
&=\sigma\left(
\sqrt{k+1}\,\pi_{k+1}(x)
+
\sqrt{k}\,\pi_{k-1}(x)
\right).
\end{align}
Thus we have
\begin{align}\label{eq:xpi}
x \pi_k(x)
=\sigma \sqrt{k+1}\,\pi_{k+1}(x)
+\sigma \sqrt{k}\,\pi_{k-1}(x).
\end{align}
and
\begin{align}
x^2 \pi_k(x)
&=\sigma^2\Big[
\sqrt{(k+1)(k+2)}\,\pi_{k+2}(x)
+(2k+1)\pi_k(x)
+\sqrt{k(k-1)}\,\pi_{k-2}(x)
\Big],
\end{align}
Using these relations, we can now integrate the individual terms of~\eqref{eq:full_integral}, starting with 

\begin{align}\label{eq:sqareterm+}
&\iint_{-\infty}^\infty dE_i\, dE_j\,
\frac{E_i^2}{N (N-1)}
e^{-(E_i^2+ E_j^2)/2\sigma^2}
\sum_{k,l = 0}^{N-1} \pi_k^2(E_i)\pi_{l}^2(E_j)
\\
&=\frac{N}{N (N-1)}
\sum_{k = 0}^{N-1}
\int_{-\infty}^\infty dE_i\, E_i^2
e^{-E_i^2/2\sigma^2} \pi_k^2(E_i)
\\
&=\frac{N\sigma^2}{N (N-1)}
\sum_{k = 0}^{N-1}
\int_{-\infty}^\infty dE_i\, e^{-E_i^2/2\sigma^2}
\pi_k(E_i)
\\
&\quad\times
\Big[
\sqrt{(k+1)(k+2)}\,\pi_{k+2}(E_i)
+(2k+1)\pi_k(E_i)
+\sqrt{k(k-1)}\,\pi_{k-2}(E_i)
\Big]
\\
&=\frac{\sigma^2}{N-1}
\sum_{k=0}^{N-1}
\Big[
\underbrace{\sqrt{(k+1)(k+2)}\,\delta_{k k+2}}_{=0}
+(2k+1)\delta_{kk}
+\underbrace{\sqrt{k(k-1)}\,\delta_{k k-2}}_{=0}
\Big]
\\
&=\frac{\sigma^2}{N-1}
\sum_{k=0}^{N-1}(2k+1)
=\frac{\sigma^2 N^2}{N-1}
\end{align}

Similarly,
\begin{align}\label{eq:sqareterm-}
&\iint_{-\infty}^\infty dE_i\, dE_j\,
\frac{E_i^2}{N (N-1)}
e^{-(E_i^2+ E_j^2)/2\sigma^2}
\sum_{k,l = 0}^{N-1}
\pi_k(E_i)\pi_{l}(E_i)\pi_k(E_j)\pi_{l}(E_j)
\\
&=\frac{\sigma^2}{N(N-1)}
\sum_{k l}
\delta_{k l}
\Big[
\sqrt{(k+1)(k+2)}\,\delta_{l k+2}
+(2k+1)\delta_{l k}
+\sqrt{k(k-1)}\,\delta_{l k-2}
\Big]
\\
&=\frac{\sigma^2}{N(N-1)}\sum_{k=0}^{N-1}(2k+1)
=\frac{\sigma^2 N}{N-1},
\end{align}
where we used $\int_{-\infty}^\infty dE_j \pi_k(E_j)\pi_l(E_j)=\delta_{kl}$. For the $E_iE_j$ terms we obtain
\begin{align}\label{eq:ijterm+}
\iint_{-\infty}^\infty dE_i\, dE_j\,
\frac{E_iE_j}{N (N-1)}
e^{-(E_i^2+ E_j^2)/2\sigma^2}
\sum_{k,l = 0}^{N-1} \pi_k^2(E_i)\pi_{l}^2(E_j)
=0,
\end{align}
since both integrals are over odd functions.

Finally,
\begin{align}\label{eq:ijterm-}
&\iint_{-\infty}^\infty dE_i\, dE_j\,
\frac{2E_iE_j}{N (N-1)}
e^{-(E_i^2+ E_j^2)/2\sigma^2}
\sum_{k,l = 0}^{N-1}
\pi_k(E_i)\pi_{l}(E_i)\pi_k(E_j)\pi_{l}(E_j)
\\
&=\frac{1}{N(N-1)}
\sum_{k,l = 0}^{N-1}
2\sigma^2
\Bigg[
\int_{-\infty}^{\infty} dE_i
\big(
\sqrt{k+1}\pi_{k+1}(E_i)
+\sqrt{k}\pi_{k-1}(E_i)
\big)
\pi_{l}(E_i)
\\
&\qquad\times
\int_{-\infty}^{\infty} dE_j
\big(
\sqrt{k+1}\pi_{k+1}(E_j)
+\sqrt{k}\pi_{k-1}(E_j)
\big)
\pi_{l}(E_j)
\Bigg]
\\
&=\frac{2\sigma^2}{N(N-1)}
\sum_{k,l =0}^{N-1}
\left(
\sqrt{k+1}\delta_{k+1,l}
+\sqrt{k}\delta_{k-1,l}
\right)^2=\frac{2\sigma^2}{N(N-1)}
\sum_{k,l =0}^{N-1}
\left(
(k+1)\delta_{k+1,l}
+k\delta_{k-1,l}
\right)
\\
&=\frac{2\sigma^2}{N(N-1)}
\sum_{k=1}^{N-1}2k
=\frac{4\sigma^2}{N(N-1)}
\left(\frac{N-1}{2}\right)N
=2\sigma^2.
\end{align}

Thus, using summing the terms we acquired from~\eqref{eq:sqareterm+},\eqref{eq:sqareterm-} and\eqref{eq:ijterm-} we get
\begin{align}
    &\int_{-\infty}^{\infty}dE_j\int_{-\infty}^{\infty}dE_i \dfrac{(E_i^2 +E_j^2 -2E_iE_j)}{N(N-1)}e^{-\frac{1}{2\sigma^2}(E_i^2+E_j^2)}\\
    &\times\left(\sum_{k,l=0}^{N-1}\pi_k^2(E_i)\pi_{l}^2(E_j)-\sum_{k,l=0}^{N-1}\pi_k(E_i)\pi_{l}(E_i)\pi_k(E_j)\pi_{l}(E_j) \right)\, \\
    &= \sigma^2[ \dfrac{2N^2}{N-1} - \dfrac{2N}{N-1} +2]=\dfrac{2\sigma^2}{N-1}(N^2-1)=2\sigma^2(N+1)
\end{align}
giving us the final result
\begin{align}\label{eq:averagesigmaG2}
\langle \sigma_G^2\rangle_{\mathrm{GUE}}
=2\sigma^2(N+1)
\end{align}
and thus the equilibration time
\begin{align}
    T_\mathrm{eq}^\mathrm{GUE}\approx\dfrac{c}{2\sigma} \dfrac{\pi}{\sqrt{2(N+1)}}
\end{align}
where the proportionally factor $c$ can be calculated numerically, which we do explicitly in App.~\ref{app:unimod-constant}.

%%%%%%%%%%%%%%%%%%%%%%%%%%%%%%%%%%%%%%%%%%%%%%%%%%%%%%%%%%%%%%%
\section{Approximation of the average}\label{app:InverseOfAverage}
In this appendix we aim to show that, for the GUE, it is valid to approximate the average of the inverse square root energy gap dispersion $\sigma_G$ as the inverse square root of its average, \emph{i.e.} that 
\begin{align}\label{eq:average_approx}
    \left\langle\frac{1}{\sqrt{\sigma_G^2}}\right\rangle_\mathrm{GUE} \approx \frac{1}{\sqrt{\left\langle\sigma_G^2\right\rangle_\mathrm{GUE}}}.
\end{align}
From the Taylor expansion of the moments of a random variable $X$, the delta method relation of~\eqref{eq:deltamethod} follows (restated here for convenience)
\begin{align}\label{eq:invaveragerelation}
    \left\langle\frac{1}{\sqrt{X}}\right\rangle \approx \frac{1}{\sqrt{\left\langle X\right\rangle_\mathrm{GUE}}}\left(1+\frac{3}{8} \frac{\operatorname{Var}\left(X\right)}{\left\langle X\right\rangle_\mathrm{GUE}^2}\right).
\end{align}
Thus if we can show that
\begin{align}
    \frac{\operatorname{Var}\left(\sigma_G^2\right)}{\left\langle\sigma_G^2\right\rangle^2_\mathrm{GUE}} \rightarrow const
\end{align}
for large $N$, we can approximate the equilibration time as $T_\mathrm{eq}\approx \tfrac{c \pi}{2\sqrt{\langle \sigma_G^2\rangle_\mathrm{GUE}}}$. 
For this we need to calculate
\begin{align}\label{eq:var4momentrelation}
    \dfrac{\operatorname{Var(\sigma_G^2)}}{\langle \sigma_G^2\rangle_\mathrm{GUE}^2}=\dfrac{\langle \sigma_G^4\rangle_\mathrm{GUE}}{\langle \sigma_G^2\rangle_\mathrm{GUE}^2}-1
\end{align}
we already know $\langle \sigma_G^2\rangle_\mathrm{GUE}$ from~\eqref{eq:averagesigmaG2}, so we only have to calculate the numerator. First we note that 
\begin{align}
    \sigma_G^{4}=\left(\sum_\alpha q_\alpha G_\alpha^2\right)^2 \leq\left(\sum_\alpha q_\alpha^2\right)\left(\sum_\alpha G_\alpha^4\right) \leq \sum_\alpha G_\alpha^4
\end{align}
where we used the (weighted) Cauchy-Schwarz inequality and the fact that $q_\alpha \geq 0$ and $\sum_\alpha q_\alpha =1$. Then
\begin{align}\label{eq:4integral}
& \left\langle\sigma_G^4\right\rangle \leq\left\langle\sum_\alpha G_\alpha^4\right\rangle=\left\langle\sum_{i j}\left(E_i-E_j\right)^4\right\rangle \\
& =\iint_{-\infty}^{\infty}  dE_i d E_j\left(E_i^4+E_j^4+6 E_i^2 E_j^2-4\left(E_i^3 E_j+E_i E_j^3\right)\right) P\left(E_i, E_j\right)
\end{align}
where, as before
\begin{align}
\label{eq:jointprob}
    P(E_i,E_j)= \dfrac{1}{N(N-1)}(K_N(E_i,E_i)K_N(E_j,E_j)-K^2_N(E_i,E_j))\,,
\end{align}
and the kernel $K_N(x,y)$ is given as 
\begin{align}
    K_N(x,y)=e^{-\frac{1}{2}(V(x)+V(y))}\sum_{j=0}^{N-1}\pi_j(x) \, \pi_j(y),
\end{align}
With this we can calculate all terms individually in a similar fashion as for the main result~\eqref{eq:averagesigmaG2}. Again it will be useful to use the relation of~\eqref{eq:xpi} to calculate higher powers of $x$:
\begin{align}
x^3 \pi_k(x)&=\sigma^3 \left(\sqrt{(k+1)(k+2)(k+3)} \pi_{k+3}(x)\right. \\
&\quad +3(k+1) \sqrt{k+1} \pi_{k+1}(x) \\
&\quad +3 k \sqrt{k} \pi_{k-1}(x) \\
&\quad \left.+\sqrt{k(k-1)(k-2)} \pi_{k-3}(x)\right) .
\\[1em]
x^4 \pi_k(x)
&=\sigma^4\Big(
\sqrt{(k+1)(k+2)(k+3)(k+4)}\,\pi_{k+4}(x)
\\
&\quad
+(4k+6)\sqrt{(k+1)(k+2)}\,\pi_{k+2}(x)
\\
&\quad
+(6k^2+6k+3)\pi_k(x)
+(4k-2)\sqrt{k(k-1)}\,\pi_{k-2}(x)
\\
&\quad
+\sqrt{k(k-1)(k-2)(k-3)}\,\pi_{k-4}(x)
\Big).
\end{align}

Starting with the $E_i^4$ term of~\eqref{eq:4integral} we get 
\begin{align}
&\iint_{-\infty}^\infty dE_i\, dE_j\,
\frac{E_i^4}{N (N-1)}
e^{-(E_i^2+ E_j^2)/2\sigma^2}
\sum_{k,l = 0}^{N-1} \pi_k^2(E_i)\pi_{l}^2(E_j)
\\
&=\frac{N \sigma^4}{N(N-1)} \sum_{k=0}^{N-1} \int_{-\infty}^{\infty} d E_i e^{-E_i^2 / 2 \sigma^2} \pi_k\left(E_i\right)\left[\sqrt{(k+1)(k+2)(k+3)(k+4)} \pi_{k+4}\left(E_i\right)\right.\\
&+\sqrt{(k+1)(k+2)}(4 k+6) \pi_{k+2}\left(E_i\right)+\left(6 k^2+6 k+3\right) \pi_k\left(E_i\right)\\
&+\left.\sqrt{k(k-1)}(4 k-2) \pi_{k-2}\left(E_i\right)+\sqrt{k(k-1)(k-2)(k-3)} \pi_{k-4}\left(E_i\right)\right]\\
&\begin{aligned}
& =\frac{\sigma^4}{(N-1)} \sum_{k=0}^{N-1}\left[\sqrt{(k+1)(k+2)(k+3)(k+4)} \delta_{k k+4}+\sqrt{(k+1)(k+2)}(4 k+6) \delta_{k k+2}\right. \\
& + \left.\left(6 k^2+6 k+3\right) \delta_{kk}+\sqrt{k(k-1)}(4 k-2) \delta_{k k-2}+\sqrt{k(k-1)(k-2)(k-3)} \delta_{k k-4}\right] \\
& =\frac{\sigma^4}{N-1} \sum_{k=0}^{N-1} 6 k^2+6 k+3=\frac{\sigma^4 N\left(1+2N^2\right)}{N-1},
\end{aligned}
\end{align}
and 
\begin{align}
&\frac{1}{N(N-1)}
\sum_{k,l=0}^{N-1}
\iint_{-\infty}^{\infty} dE_idE_j
E_i^{4}\,
e^{-(E_i^2+E_j^2)/2\sigma^2}\,
\pi_k(E_i)\pi_l(E_i)\pi_k(E_j)\pi_l(E_j)
\\
&=
\frac{\sigma^4}{N(N-1)}
\sum_{k=0}^{N-1}
\int_{-\infty}^{\infty} dE_i\,
e^{-E_i^2/2\sigma^2}\,
\pi_k(E_i)
\Big[
\sqrt{(k+1)(k+2)(k+3)(k+4)}\,\pi_{k+4}(E_i)
\\
&\qquad\qquad
+ \sqrt{(k+1)(k+2)}(4k+6)\,\pi_{k+2}(E_i)
+ (6k^2+6k+3)\,\pi_k(E_i)
\\
&\qquad\qquad
+ \sqrt{k(k-1)}(4k-2)\,\pi_{k-2}(E_i)
+ \sqrt{k(k-1)(k-2)(k-3)}\,\pi_{k-4}(E_i)
\Big]
\\
&=
\frac{\sigma^4}{N(N-1)}
\sum_{k=0}^{N-1}
\Big[
\sqrt{(k+1)(k+2)(k+3)(k+4)}\,\delta_{k,k+4}
\\
&\qquad
+ \sqrt{(k+1)(k+2)}(4k+6)\,\delta_{k,k+2}
+ (6k^2+6k+3)\,\delta_{k,k}
\\
&\qquad
+ \sqrt{k(k-1)}(4k-2)\,\delta_{k,k-2}
+ \sqrt{k(k-1)(k-2)(k-3)}\,\delta_{k,k-4}
\Big]
\\
&=
\frac{\sigma^4}{N(N-1)}
\sum_{k=0}^{N-1}
\left(6k^2+6k+3\right)
\\
&=\frac{\sigma^4 \left(1+2N^2\right)}{N-1}
.
\end{align}
Next we look at the terms coming from $E_i^2E_j^2$, for which we get
\begin{align}
    &\iint_{-\infty}^\infty dE_i\, dE_j\,
\frac{6E_i^2E_j^2}{N (N-1)}
e^{-(E_i^2+ E_j^2)/2\sigma^2}
\sum_{k,l = 0}^{N-1} \pi_k^2(E_i)\pi_{l}^2(E_j)\\
&=\frac{6 \sigma^4}{N(N-1)} \sum_{k=0}^{N-1}(2 k+1) \sum_{l=0}^{N-1}(2 l+1)=6\sigma^4 \frac{N^3}{N-1},
\end{align}
where we used the integral from~\eqref{eq:sqareterm+}, and
\begin{align}
    &\iint_{-\infty}^\infty dE_i\, dE_j\,
\frac{6E_i^2E_j^2}{N (N-1)}
e^{-(E_i^2+ E_j^2)/2\sigma^2}
\sum_{k,l = 0}^{N-1} \pi_k(E_i)\pi_{l}(E_i)\pi_k(E_j)\pi_{l}(E_j)=6\sigma^4\dfrac{ N^3}{N-1}
\end{align}
where we use the integral from~\eqref{eq:sqareterm-} and squared the result. These two terms thus cancel.

Finally we come to the terms $E_i^3E_j$. First, note that the term $-4\left(E_i^3 E_j+E_j E_i^3\right)\sum_{k,l=0}^{N-1} \pi_k^2(E_i)\pi_l^2(E_j)$ results in an odd function which gives 0 if integrated over all real numbers. Thus we only have to calculate
\begin{align}
    &\dfrac{4}{N(N-1)}\iint_{-\infty}^\infty dE_i\, dE_j\,
E_i^3E_j
e^{-(E_i^2+ E_j^2)/2\sigma^2}
\sum_{k,l = 0}^{N-1} \pi_k(E_i)\pi_{l}(E_i)\pi_k(E_j)\pi_{l}(E_j)\\
&=\dfrac{4}{N(N-1)} \sum_{k,l=0}^{N-1}\int_{-\infty}^\infty dE_j e^{-E_j^2 / 2 \sigma^2} \pi_l\left(E_j\right)\left[\sqrt{k+1} \pi_{k+1}\left(E_j\right)+\sqrt{k} \pi_{k-1}\left(E_j\right)\right]\\
& \times \int_{-\infty}^{\infty} d E_i e^{-E_i^2 / 2 \sigma^2} \pi_l\left(E_i\right)\left[\sqrt{(k+1)(k+2)(k+3)} \pi_{k+3}\left(E_i\right)+3(k+1) \sqrt{k+1} \pi_{k+1}\left(E_i\right) \right. \\
& \left.\quad+3 k \sqrt{k} \pi_{k-1}\left(E_i\right)+\sqrt{k(k-1)(k-2)} \pi_{k-3}\left(E_i\right)\right]\\
&=\frac{4 \sigma^4}{N(N-1)} \sum_{k ,l =0}^{N-1}\left(\sqrt{k+1} \delta_{l k+1}+\sqrt{k} \delta_{l k-1}\right) \left(\sqrt{(k+1)(k+2)(k+3)} \delta_{l k+3}+\right. \\
&\left.+3(k+1) \sqrt{k+1} \delta_{l k+1}+3 k \sqrt{k} \delta_{l k-1}+\sqrt{k(k-1)(k-2)} \delta_{l k-3}\right)=\frac{4 \sigma^4}{N(N-1)} \cdot(N-1) N(2 N-1)=4 \sigma^4(2 N-1)
\end{align}
where the second to last equality uses that only the $\delta_{l,k\pm1}$ terms survive. Combining all these terms we get the following result
\begin{align}
    \langle \sigma_G^4\rangle_\mathrm{GUE}\leq4 \frac{\sigma^4 N\left(1+2N^2\right)}{N-1}- 4 \frac{\sigma^4 \left(1+2N^2\right)}{N-1} -(-4 \sigma^4(2 N-1))=8\sigma^4 N(N+1)
\end{align}
and thus we get for the ratio
\begin{align}
    \dfrac{\langle \sigma_G^4\rangle_\mathrm{GUE}}{\langle \sigma_G^2\rangle_\mathrm{GUE}^2}\leq\dfrac{2N(N+1)}{(N+1)^2}=\dfrac{2N}{N+1}\implies \dfrac{\operatorname{Var}(\sigma_G^2)}{\langle \sigma_G^2\rangle_\mathrm{GUE}}\leq \dfrac{N-1}{N+1}\rightarrow 1
\end{align}
in the limit of large $N$. Using~\eqref{eq:var4momentrelation} this means that that~\eqref{eq:invaveragerelation} does not diverge with $N$ and is bounded as
\begin{align}
    0 \leq \frac{\operatorname{Var}\left(\sigma_G^2\right)}{\left\langle\sigma_G^2\right\rangle^2_\mathrm{GUE}} \leq \frac{N-1}{N+1} \underset{N \rightarrow \infty}{\longrightarrow} 1 .
\end{align}
and thus
\begin{align}
     \left\langle\frac{1}{\sqrt{\sigma_G^2}}\right\rangle_\mathrm{GUE} \approx \frac{c}{\sqrt{\left\langle\sigma_G^2\right\rangle_\mathrm{GUE}}}
\end{align}
where $c\approx\left( 1+\tfrac{3}{8}\tfrac{\operatorname{Var}\left(X\right)}{\left\langle X\right\rangle_\mathrm{GUE}^2}\right)\underset{N\rightarrow\infty}{\longrightarrow}1.375$ is a constant, thus showing the validity of~\eqref{eq:average_approx}.

\section{Numerical demonstration of the convergence of $c$ for large $N$}
\label{app:unimod-constant}
In App.~\ref{app:InverseOfAverage}, we showed that the average of the inverse square root of the gap dispersion can be approximated by the inverse square root of its average, up to a constant factor $c$. Here we verify this numerically and evaluate $c$ for large~$N$. 

\begin{figure}[htbp]
    \centering
    \includegraphics[width=0.5\linewidth]{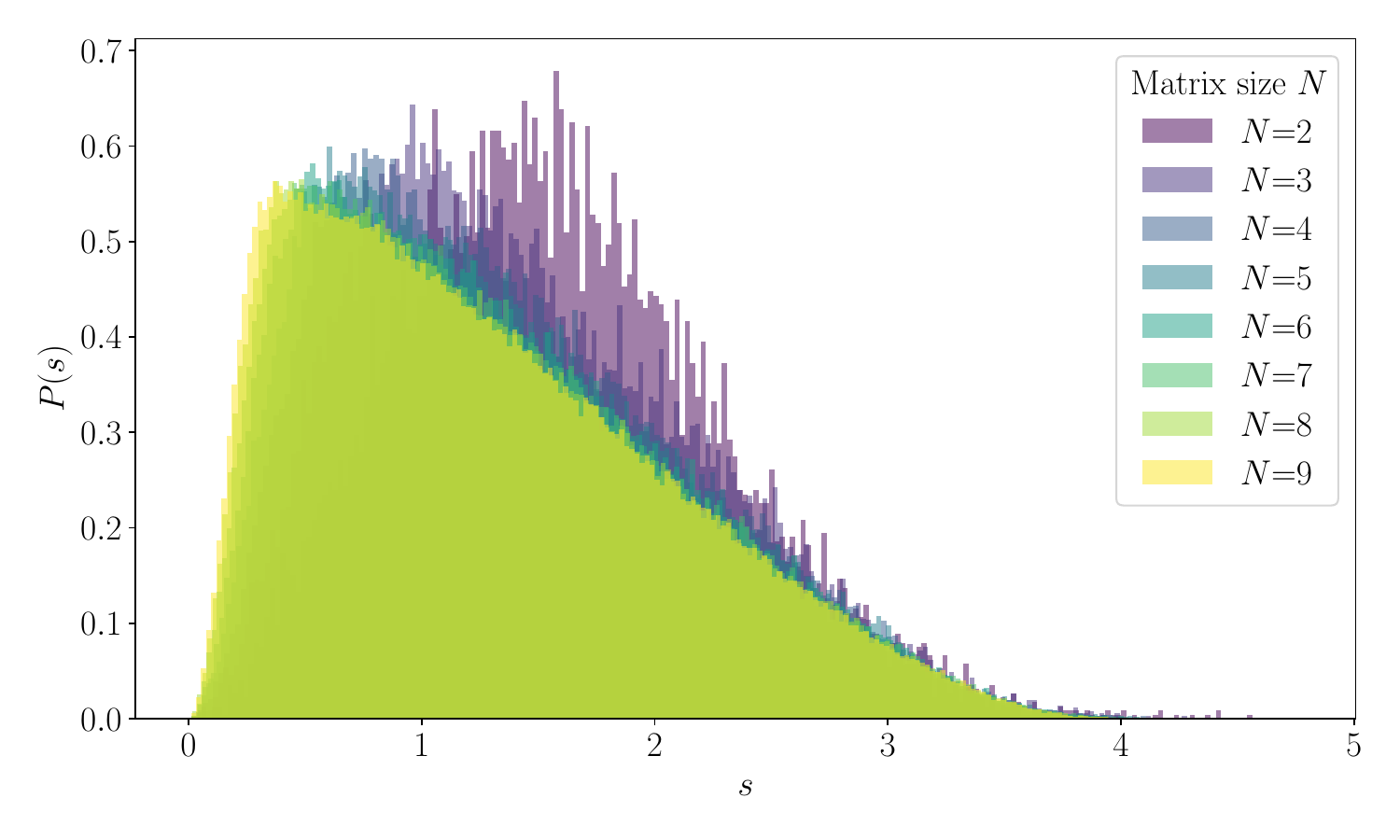}
    \caption{Plot of average spacing distribution $P(s)$ as a function of the gap $s=\vert E_i - E_j\vert$ for different values of the system size~$N$. The standard deviation of the sampled GUE Hamiltonians is set to $\sigma = 1/\sqrt{N}$. For large $N$ this distribution is unimodal, thus providing further justification for the equilibration time~\eqref{eq:EquilibrationTimeHeuristic}.}
    \label{fig:P(s)numerics}
\end{figure}

To build intuition for the level spacing distribution, we first illustrate it by generating $10^5$ sample matrices drawn from the GUE according to~\eqref{eq:GUEelements}. We set $\sigma = 1/\sqrt{N}$ to ensure a finite spectral support (a standard rescaling used, for instance, in the derivation of the Wigner semicircle law), and vary the matrix dimension from $2$ to $10$. Fig.~\ref{fig:P(s)numerics} shows a histogram of the relative frequencies of gaps of size $s$. The distribution is unimodal and short-tailed, and distributions at larger dimensions become increasingly similar, overlapping substantially — a manifestation of self-averaging in GUE ensembles. This distribution determines the the probability of obtaining a gap of size $s$ when drawing from the GUE. It replaces the gap relevances in the calculation of the average equilibration time (see App.~\ref{app:Bounding the equilibration time of Gaussian Unitary Ensembles}). Note that, in contrast to the Wigner surmise, the gaps $s$ here range over all pairs of eigenvalues, not just nearest neighbours.

Next we numerically demonstrate that $c$ tends towards a constant value for large dimensions $N$ and calculate its value. To this end, we numerically evaluate the dynamics of GUE Hamiltonians embedding a notion of locality in the Hamiltonian by considering only $N = 2^L$. This allows us to understand the GUE sampled matrix $H_{\text{GUE}}$ as the Hamiltonian of a system composed of $L$ spins $1/2$ with all-to-all interactions. Given this notion of locality, we analyse the equilibration properties of the system by evolving a local observable $A$ (specifically the sum of local spin operators) and analysing its relaxation timescale, where $A$ is given as 
\begin{equation}
    A = \sum_{i=1}^{L} \sigma_z^{(i)}\, ,
\end{equation}
and, where $\sigma_z^{(i)} = \mathbb{I}_{1:i-1} \otimes \sigma_z \otimes \mathbb{I}_{i+1:L}$, and $\mathbb{I}_{x:y}$ denotes the identity operator on the subsystems labeled from $x$ to $y$.
\par 

\begin{figure}[htbp!]
        \centering
        \includegraphics[width=0.5\linewidth]{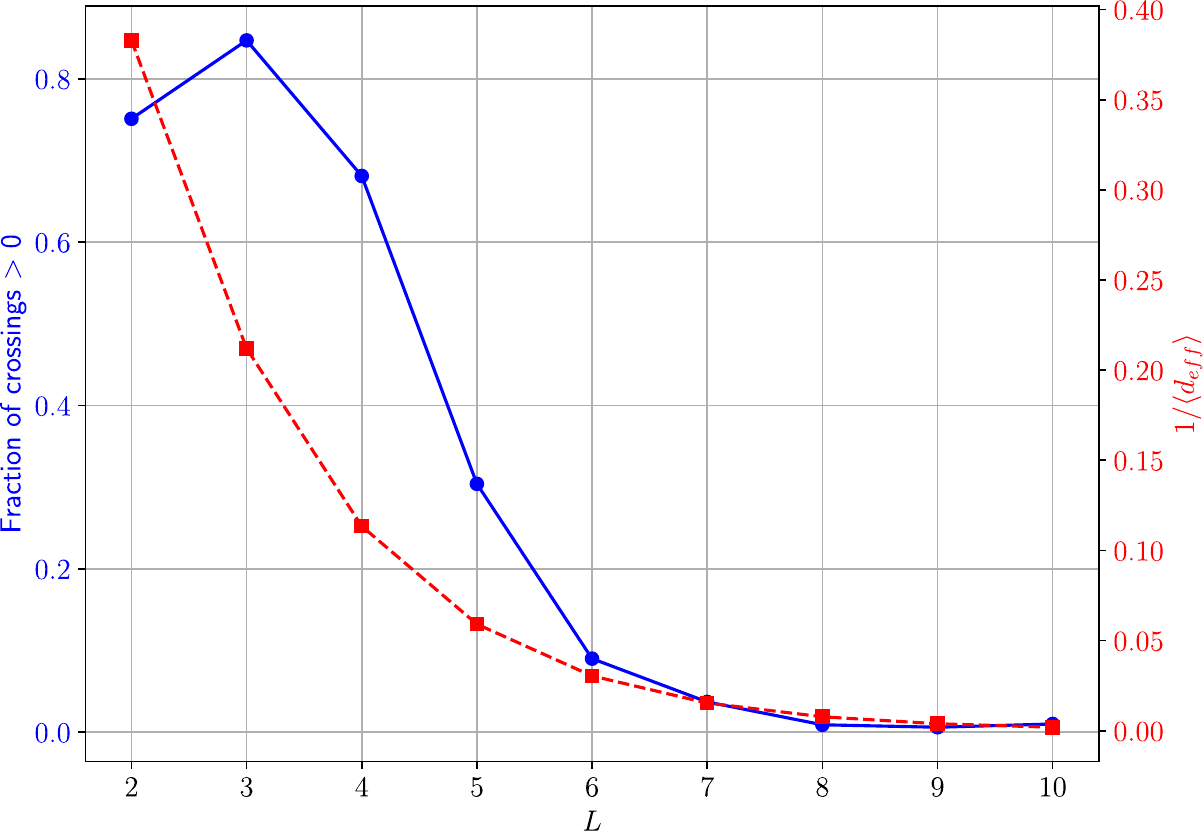}
        \label{fig:fig1appNum}
     \caption{
    The blue line corresponds to the fraction of crossings of the bound in~\eqref{eq:boundavgtimsig} for individual realization of $H_{\text{GUE}}$ as a function of $L$, where $N=2^L$; the red dashed line corresponds to the inverse of the average effective dimension $\tfrac{1}{d_\mathrm{eff}}$ as a function of $L$. For large $L$, both, the crossings and the effective dimension vanish, indicating that the system equilibrates and stays close to equilibrium, not only on average, but even for individual realisations. }
\end{figure}

We simulate the evolution of $\langle A\rangle_{\rho(t)}$ to $L=10$, and report $\langle g(t)^2\rangle_{\text{ens}}$ in Fig. \ref{fig:fig1appNum}, the averaged signal over an ensemble of different independent realisation of $H_{\text{GUE}}$. 
We recall that the latter is bounded by the effective dimension as \cite{reimann2008foundation,short2011}
\begin{equation}
    \langle \vert g(t)\vert^2 \rangle_{\infty} \leq \frac{1}{d_{\text{eff}}}\, .
    \label{eq:boundavgtimsig}
\end{equation}

To identify the time at which the system has effectively equilibrated, we record the \emph{first-crossing time} $t^{(k)}_{\mathrm{fc}}$, defined as the first time the signal $g^2_k(t)$, sampled from an individual Hamiltonian $H^{(k)}$ drawn from the GUE, drops below its infinite-time average. The mean first-crossing time over all samples,
\begin{align}\label{eq:proxy}
    t_\mathrm{fc}=\tfrac{1}{M}\sum_{k=1}^M t_\mathrm{fc}^{(k)},
\end{align}
serves as a proxy for the average equilibration time, since deviations from this value become exponentially suppressed with increasing system dimension (see~\eqref{eq:reimann}). Here $k$ indexes individual samples from the GUE and $M$ denotes the total number of samples.

In Fig.~\ref{fig:fig1appNum} we report the average number of times the signal fluctuates above this bound, together with the inverse of the average effective dimension. As $L$ increases, the inverse effective dimension decreases (tightening the bound in~\eqref{eq:boundavgtimsig}) and simultaneously, the number of trajectories that exceed this bound also decreases, as one would expect. We recall that this bound is always satisfied by the infinite-time average of $g(t)$, however, transient fluctuations may allow averages over shorter time windows to temporarily exceed it. The fact that the bound vanishes while the number of trajectories overcoming it decreases simultaneously thus provides a clear signature that the system has effectively equilibrated over most time intervals. Therefore, for sufficiently large dimensions, the first-crossing time serves as a reliable proxy for the equilibration time, \emph{i.e.},
\begin{align}
    t_\mathrm{fc}\approx T_\mathrm{eq}^\mathrm{GUE}.
\end{align}
We are now ready to discuss the scaling of the proportionality constant $c$ as a function of $L$. Using~\eqref{eq:Teq} and substituting the average equilibration time with the proxy from~\eqref{eq:proxy}, we can evaluate the numerical proportionality constant for a matrix of size $N$, $c_\mathrm{Num}(N)$ as
\begin{equation}
    c_\mathrm{Num}(N) \approx t_\mathrm{fc}\dfrac{2\sqrt{2\sigma^2(N+1)}}{\pi}\, .
\end{equation}
 We plot its values in Fig. \ref{fig:fit_c_vs_L}, where we fit ${c_\mathrm{Num}(N)}$ with respect to a shifted exponential function ${ y_\mathrm{fit}(L) = A e^{-B*L} + c_\mathrm{Num}}$. The fit yields to the following values: ${A=  -0.94 ± 0.03}$, ${B =0.44 ± 0.02}$ and ${c_\mathrm{Num}= 1.368 \pm 0.004}$, where ${c_\mathrm{Num}:=c_\mathrm{Num}(N\rightarrow\infty)}$. Furthermore, to evaluate the goodness of the fit, we compute the root mean square error (RMSE), obtaining the value ${\textup{RMSE} = 0.005916}$. Fig.~\ref{fig:fit_c_vs_L} shows us that the constant $c_\mathrm{Num}(N)$, while changing for small dimensions, approaches a finite constant value for large $N$. 
\end{document}